\documentclass[12pt,a4paper]{article}

\usepackage{amsthm}
\usepackage{amssymb}
\usepackage{amsmath}
\usepackage{graphicx}
\usepackage{url}

\newcommand{\E}{{\mathbb E}}
\newcommand{\SNR}{{\mathrm{SNR}}}

\begin{document}

\title{The statistical properties of RCTs \\and a proposal for shrinkage}
\author{Erik van Zwet \and Simon Schwab \and Stephen Senn}

\author{Erik van Zwet\footnote{Department of Biomedical Data Sciences, Leiden University Medical Center.} \qquad Simon Schwab\footnote{1. Center for Reproducible Science (CRS), University of Zurich, Switzerland. 2. Epidemiology, Biostatistics and Prevention Institute (EPBI), University of Zurich, Switzerland.
} \qquad Stephen Senn \footnote{Statistical consultant.}}
\maketitle

\begin{abstract}
We abstract the concept of a randomized controlled trial (RCT) as a triple $(\beta,b,s)$, where $\beta$ is the primary efficacy parameter, $b$ the estimate and $s$ the standard error ($s>0$). The parameter $\beta$ is either a difference of means, a log odds ratio or a log hazard ratio. If we assume that $b$ is unbiased and normally distributed, then we can estimate the full joint distribution of $(\beta,b,s)$ from a sample of pairs $(b_i,s_i)$.  We have collected 23,747 such pairs from the Cochrane database to do so. Here, we report the estimated distribution of the signal-to-noise ratio $\beta/s$ and the achieved power. We estimate the median achieved power to be 0.13. We also consider the exaggeration ratio which is the factor by which the magnitude of $\beta$ is overestimated. We find that if the estimate is just significant at the 5\% level, we would expect it to overestimate the true effect by a factor of 1.7. This exaggeration is sometimes referred to as the winner's curse and it is undoubtedly to a considerable extent responsible for disappointing replication results. For this reason, we believe it is important to shrink the unbiased estimator, and we propose a method for doing so.
\end{abstract}


\section{Introduction}
It is nearly three quarters of a century since what is generally regarded as the first modern randomized clinical trial, the UK Medical Research Council study of the effectiveness of streptomycin in tuberculosis \cite{medical1948streptomycin}. Since then, tens of thousands of randomized controlled trials (RCT) have been conducted. The purpose of this paper is to study this wealth of information, and to try to learn from it.

We have collected the results of more than 20,000 RCTs from the Cochrane Database of Systematic Reviews (CDSR), which is the leading journal and database for systematic reviews in health care. These data allow us to determine the broad statistical properties of RCTs. In particular, we are able to estimate the distribution of the achieved power across all RCTs. We find that the achieved power is often quite low. Low statistical power has been noticed before in specific domains of biomedical research, see for instance \cite{button2013power, dumas2017low}. 

We hasten to say that the fact that achieved power is typically low does not imply that the usual sample size calculations aiming for 80\% or 90\% power are wrong. The goal of such calculations is to guarantee high power against a particular alternative that is considered to be of clinical interest. The fact that high power is often not achieved is merely an indication that treatments often do not provide the benefit that was hoped for.

Low power has important implications for the interpretation of the results of a given trial. It is well known that conditional on statistical significance, the estimate of the effect size is positively biased. This bias is sometimes called the ``winner's curse'', and it is especially large when the power is low \cite{ioannidis2008most, gelman2014beyond, vanZwetCator}. With the Cochrane data we can quantify the bias quite precisely. Finally, we can determine the coverage of the usual 95\% confidence interval, conditional on the observed result of the trial.

Overestimation of the effect size and undercoverage of the confidence interval are obviously serious issues which are part of the explanation for the phenomenon of poor replication  \cite{ioannidis2005contradicted, ioannidis2008most, button2013power, button2013empirical}. In section \ref{shrinkage} we propose an empirical Bayesian shrinkage estimator to mitigate these problems. We conclude this short paper with a discussion.

\section{Statistical analysis of a collection of RCTs}
In this paper, we abstract the concept of a randomized controlled trial (RCT) as a triple $(\beta,b,s)$, where $\beta$ is the primary efficacy parameter, $b$ the estimate and $s$ the standard error ($s>0$). The parameter $\beta$ is either a difference of means, a log odds ratio or a log hazard ratio.  We will ignore small sample issues by assuming that $b$ is a normally distributed, unbiased estimator of $\beta$ with known standard error $s$. In other words,  we will make the following assumption.

\medskip \noindent {\bf Assumption 1} Conditionally on $\beta$ and $s$, $b$ is normally distributed with mean $\beta$ and standard deviation $s$. 

\medskip \noindent 
While this assumption may appear to be overly simplistic, we emphasize that inference based on Wald type confidence intervals and associated $p$-values is very common. When the sample size is not too small, say at least 60, this is quite appropriate. 

We never observe the true effect $\beta$, but only the pair $(b,s)$. However, Assumption 1 implies that if we have a sample of such pairs then we can still estimate the complete joint distribution of $(\beta,b,s)$, which we will denote by $f(\beta,b,s)$. Indeed, it is clear that we can estimate the marginal distribution $f(s)$ of the standard error, and the conditional distribution $f(b \mid s)$ of the estimator given $s$. Conditionally on $s$, $\beta$ is the sum of $b$ and independent, normally distributed noise with mean zero and standard deviation $s$. Hence, the conditional distribution $f(\beta \mid s)$ can be obtained by deconvolution of $f(b \mid s)$. Finally, the distribution of $b$ given $\beta$ and $s$ is already given by Assumption 1. 

The Cochrane data are publicly available and we have extracted  pairs $(b_i,s_i)$ from 23,747 unique RCTs. Each pair belongs to a different study, and we have tried to obtain the estimate and standard error of the first or primary effect. For additional details about the data collection we refer to Schwab et al.\  \cite{schwab}.

Under Assumption 1 we can estimate the full joint distribution of $(\beta,b,s)$ from a sample of pairs $(b_i,s_i)$. This joint distribution has many important aspects. In particular, we will discuss the conditional (``posterior") distribution of $\beta$ given $b$ and $s$ in section \ref{shrinkage}. However, we will focus first on the joint distribution of the
signal-to-noise ratio $\SNR=\beta/s$ and the $z$-value $z = b/s$. Many interesting statistical properties, such as the power, exaggeration and coverage can be studied using only the joint distribution of $\SNR$ and $z$.

It is obvious that we can estimate the marginal  distribution of the $z$-value from the pairs $(b_i,s_i)$. We will make the following assumption.

\medskip \noindent {\bf Assumption 2} The distribution of $z=b/s$ is a finite mixture of zero-mean normal distributions.

\begin{figure}[htp] \centering{
\includegraphics[scale=0.8]{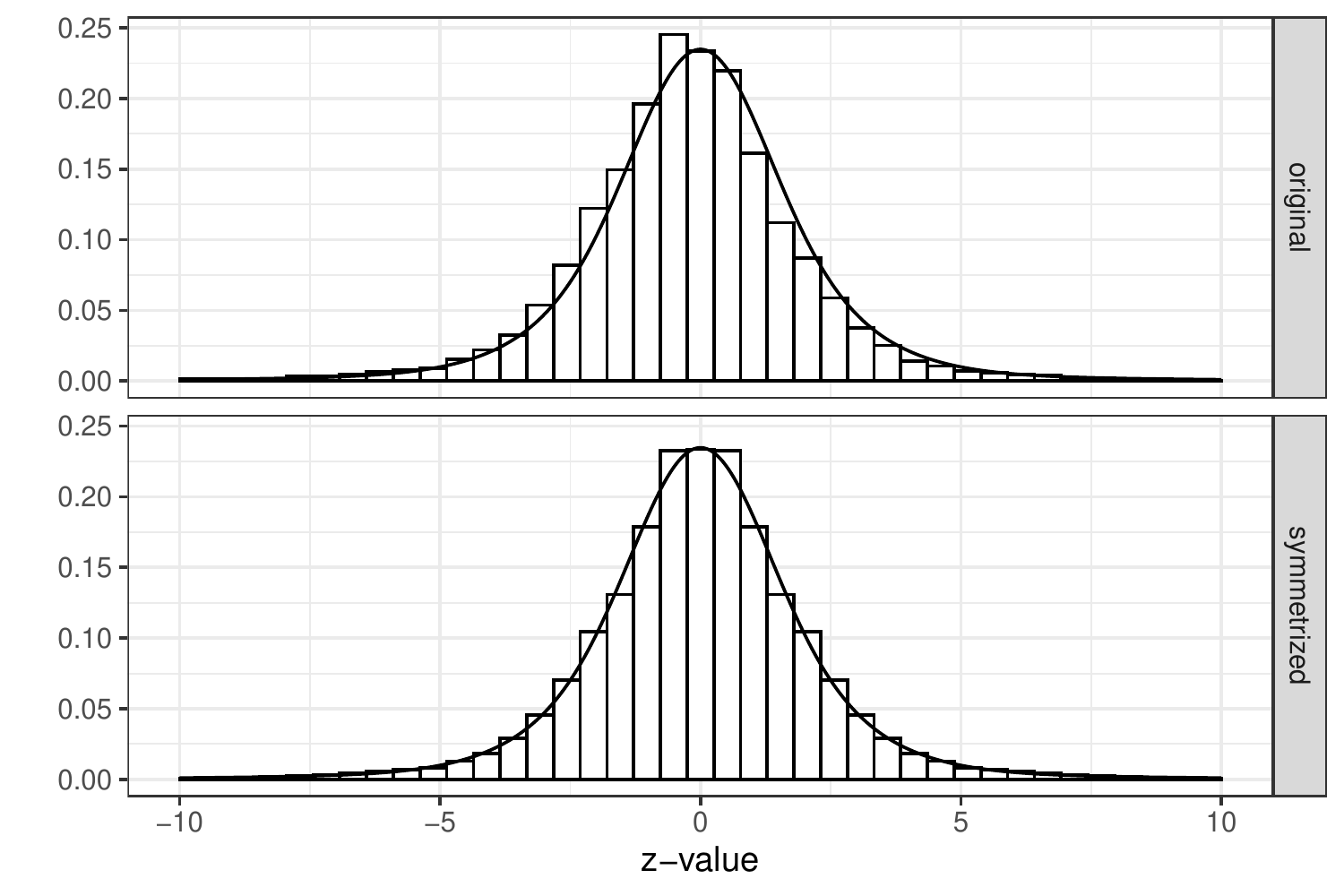}
\caption{Top panel: The histogram of the observed $z$-values together with our fit based on a mixture of 4 zero-mean normal distributions. Bottom panel: The symmetrized histogram together with the same fit}\label{fig:z}
}
\end{figure}

\medskip \noindent
We have used the R package ``flexmix'' \cite{flexmix}, which implements the EM algorithm, to estimate the normal mixture. We tried 1 up to 6 components, and found that more than 4 components made no discernible difference to the estimated distribution.  In the top panel of Figure \ref{fig:z} we show the histogram of the observed $z$-values together with our estimate based on a mixture of 4 zero-mean normals. The estimates of the mixture components are given in Table \ref{table:mix}. Our mixture fits reasonably well, as the distribution of $z$-values is roughly symmetric around zero. This symmetry was also observed by Djulbegovic et al.\ who conducted a large meta-analysis of 860 published and unpublished phase III RCTs \cite{djulbegovic2013trial}.

While the symmetry of the distribution of the $z$-values is nice, it is not so relevant for us. We are not interested in the sign of the $z$-value, for two reasons. First, the sign is determined by the arbitrary choice of comparing arm A to B or vice versa. Second, many of the results that we will present depend on the $z$-value only though its absolute value. In the bottom panel of Figure \ref{fig:z} we show the {\it symmetrized} histogram of the observed $z$-values, also with our estimate. We note the excellent fit.

By Assumption 1, the $z$-value is the sum of the signal-to-noise ratio $\SNR=\beta/s$ and independent standard normal noise. So, we can obtain the distribution of the SNR by deconvolution. Since the distribution of the $z$-value is a mixture of normal distributions, this is particularly easy; we simply subtract 1 from the variances of the components. The resulting standard deviations  are given in Table \ref{table:mix}. In Table \ref{table:power} we report the 10, 25, 50, 75 and 90 percentiles of the distribution of the absolute value of the SNR.

\bigskip

\begin{table}[ht]
\centering
\begin{tabular}{rrrrr}
  \hline
 & comp.1 & comp.2 & comp.3 & comp.4 \\ 
  \hline
proportions & 0.32 & 0.31 & 0.30 & 0.07 \\ 
  std. dev. of the $z$-value & 1.19 & 1.71 & 2.40 & 5.65 \\ 
  std. dev. of the SNR & 0.64 & 1.38 & 2.18 & 5.56 \\ 
   \hline
\end{tabular}
\caption{Estimated 4-part normal mixture distributions of the $z$-value and the SNR.}
\label{table:mix}
\end{table}

\begin{table}[h]
\centering
\begin{tabular}{rrrrrr}
  \hline
 & Q10 & Q25 & Q50 & Q75 & Q90 \\ 
  \hline
$|\SNR|$ & 0.14 & 0.37 & 0.84 & 1.72 & 3.01 \\ 
  Power & 0.05 & 0.07 & 0.13 & 0.41 & 0.85 \\ 
  Exaggeration ratio & 16.01 & 6.31 & 2.90 & 1.55 & 1.09 \\ 
   \hline
\end{tabular}
\caption{Estimated quantiles of the absolute value of the signal-to-noise ratio, the power and the exaggeration ratio.}
\label{table:power}
\end{table}

\section{Power, Exaggeration and Coverage}
In this section we quantify the achieved power, exaggeration (i.e.\ overestimation of the magnitude of $\beta$), and coverage relative to the population of RCTs from the Cochrane database.

\subsection{Power}
\begin{figure}[htp] \centering{
\includegraphics[scale=0.8]{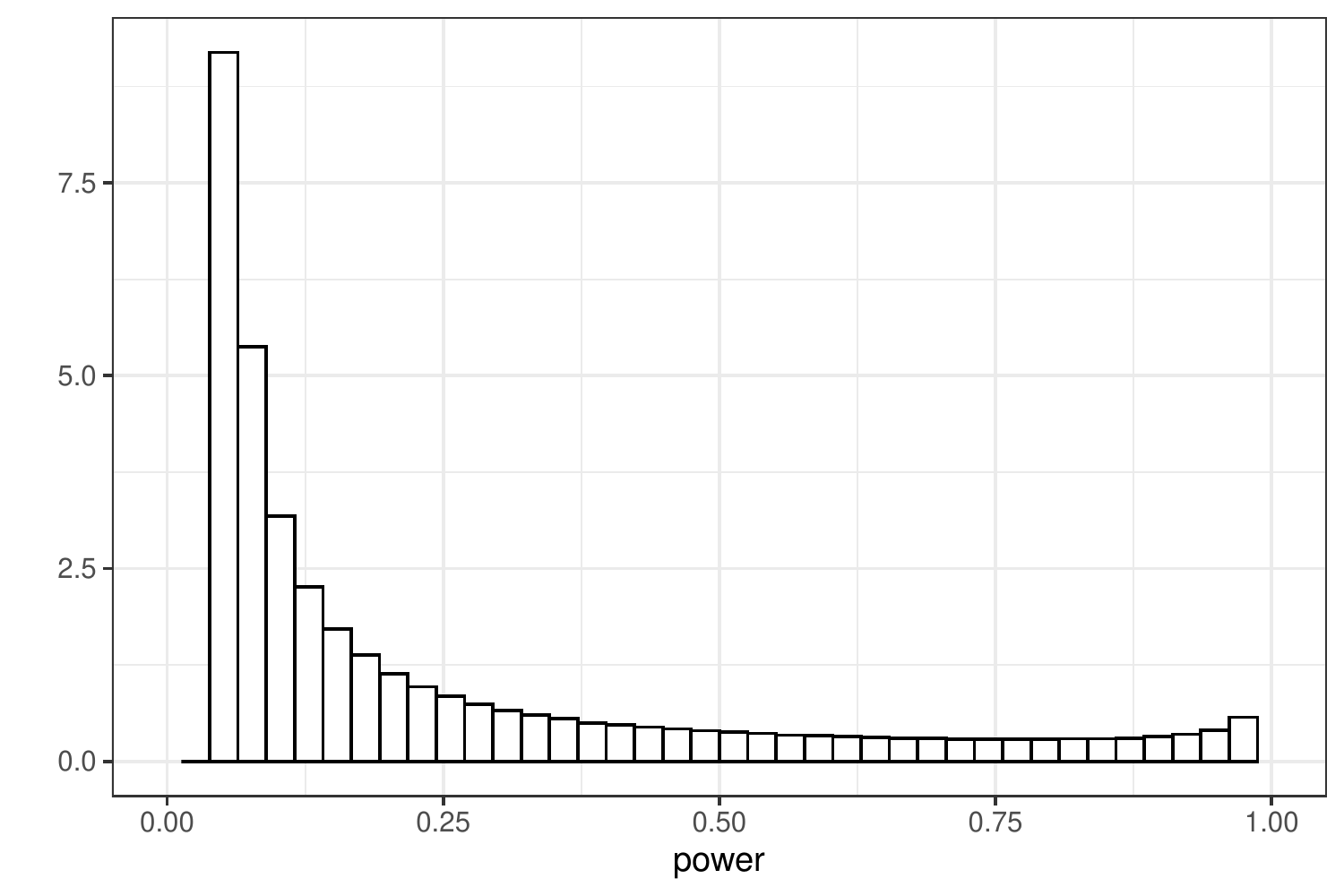}
\caption{Histogram of a sample of size $10^6$ from the estimated distribution of the power.}\label{fig:power}
}
\end{figure}

The achieved power of the two-sided Wald test of $H_0 : \beta = 0$ at level 5\% depends on the SNR and is given by
\begin{equation}\label{power}
\text{power} = \Phi(-1.96 - \SNR) + 1 - \Phi(1.96 - \SNR),
\end{equation}
where $\Phi$ is the cumulative distribution function of the standard normal distribution. The power is an even function, so it depends on the SNR only through its absolute value. In fact, the power is a strictly increasing function of absolute value of the SNR. Using formula (\ref{power}), it is easy to transform a sample from the distribution of the SNR into a sample from the distribution of the power. We generated such a sample of size $10^6$ and show the histogram  in Figure \ref{fig:power}.  In Table \ref{table:power} we report a number of quantiles of the distribution of the power. For example, the median of the power is 13\%. The average power is 28\%. O'Hagan et al. \cite{o2005assurance} discuss the importance of average power to which they refer as the ``assurance''.

Most RCTs are designed to have 80\% or 90\% power against an alternative that is considered to be of clinical interest. As we can see from Figure \ref{fig:power} and Table \ref{table:power}, the {\em achieved} power is usually much lower than 80\%. Indeed, we estimate that about 88\% of RCTs have power less than 80\%. However, this does not mean that most sample size calculations are mistaken. It merely indicates that many treatments do not have the effect that was considered to be of clinical interest. Finding good treatments is not easy! Other factors may also contribute to low power, such as difficulties with subject recruitment or larger between-subject variation than anticipated. In any case, the fact remains that the actual power in the majority RCTs is quite low, and this has consequences for our inferences.

\subsection{Exaggeration}
An important quantity is the ratio $R=|b|/|\beta|.$ Its expectation $\E(R \mid \beta,s)$ could be called the relative bias of the magnitude. In the absence of bias, it is equal to one. However, the fact that $b$ is unbiased for $\beta$ implies that $|b|$ is positively biased for $|\beta|$ (Jensen's inequality). So the unbiasedness of $b$ implies that $\E(R \mid \beta,s)$ is always greater than one. 

Gelman and Carlin \cite{gelman2014beyond} define the  ``expected type M error'' or ``exaggeration ratio'' as 
\begin{equation}
\E(R \mid \beta, s, |b|/s>1.96).
\end{equation}
It represents the factor by which the magnitude of the effect may be expected to be overestimated when we condition on significance at the 5\% level (two-sided). This exaggeration is sometimes referred to as the winner's curse. Undoubtedly, the winner's curse is to a considerable extent responsible for disappointing replication results. For this reason, we believe it is important to apply some regularization or ``shrinkage'' to the unbiased estimator. For a slightly different view on shrinkage, see \cite{senn2008transposed}. 

\begin{figure}[htp] \centering{
\includegraphics[scale=0.8]{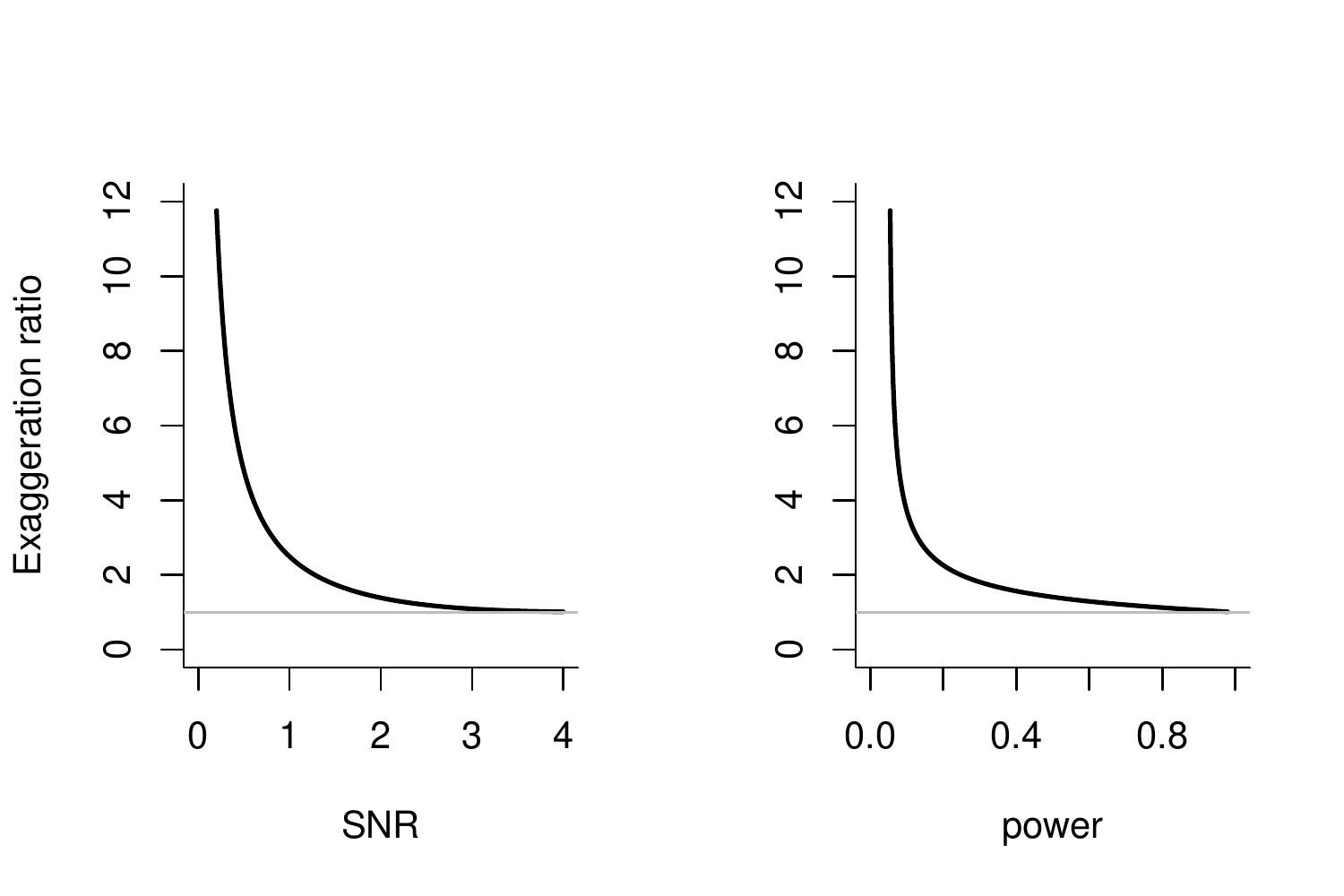}
\caption{The exaggeration ratio (expected type M error) as a function of the SNR and the power.}\label{fig:m}
}
\end{figure}

It turns out that the exaggeration ratio depends on $\beta$ and $s$ only through the absolute value of the SNR.
In fact, van Zwet and Cator \cite{vanZwetCator} show that the exaggeration ratio is decreasing in $|\SNR|$. Since the power is a strictly increasing function of  $|\SNR|$, it follows the exaggeration ratio is also a decreasing function of the power, see also \cite{ioannidis2008most} and \cite{gelman2014beyond}. In Figure \ref{fig:m} we show the exaggeration ratio as a function of $|\SNR|$ and as a function of the power. 

From Table \ref{table:power} we know that the median power of the RCTs in our study is 0.13. From Figure \ref{fig:m} we see that if the power is 0.13, then the exaggeration ratio is about 2.9; see also Table \ref{table:power}. This means that if a study with median power reaches significance, we would expect the magnitude of the effect to be overestimated by a factor of 2.9. 

Of course, in practice we do not know $\beta$ so we know neither the power nor the exaggeration ratio. However, as we have discussed, Assumption 1 makes it possible to estimate the complete joint distribution of $(\beta,b,s)$. So, in particular it is possible to estimate the conditional distribution of $R$ given $b$ and $s$. 

For simplicity, however, we will first consider the conditional distribution of $R$ given {\em only} $z=b/s$. We will return to this simplification in section \ref{shrinkage}. We have
\begin{equation}
R=\frac{|b|}{|\beta|} = \frac{|b|/s}{|\beta|/s} = \frac{|z|}{|\SNR|},
\end{equation}
so we can calculate the conditional distribution of $R$ given $z$ if we have the conditional distribution of the $\SNR$ given $z$. The marginal distributions of the $z$-value and the SNR are mixtures of normal distributions which are specified in Table \ref{table:mix}. Moreover, the $z$-value is the sum of the SNR and independent standard normal noise. Therefore, it is quite easy to calculate the conditional distribution of the SNR given $z$. 

\begin{figure}[htp] \centering{
\includegraphics[scale=0.8]{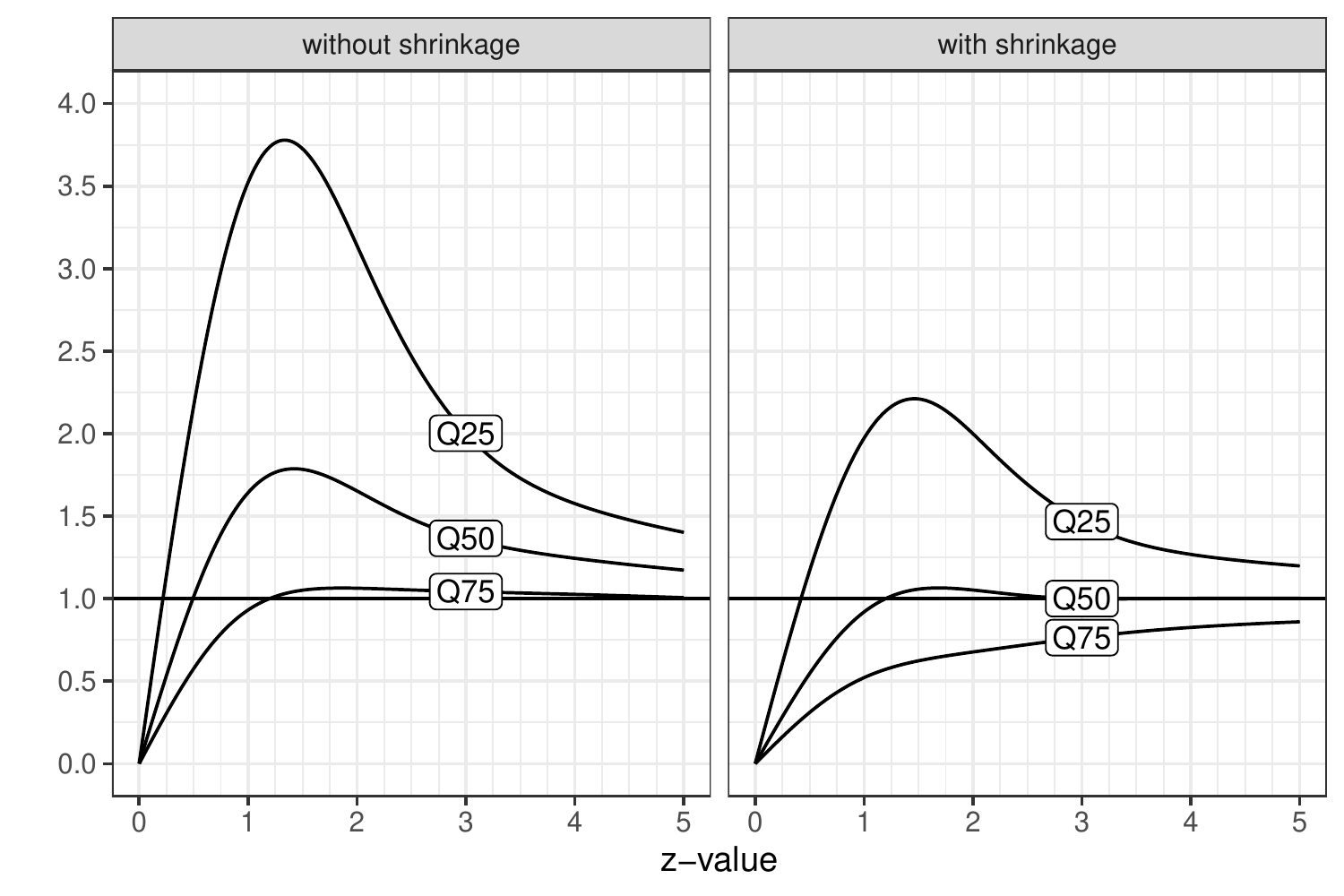}
\caption{Left panel: The distribution of the exaggeration ratio $R=|b|/|\beta|$ conditional on the $z$-value, represented by its three quartiles. Right panel: The distribution of the ratio $|\hat \beta|/|\beta|$ conditional on the $z$-value}\label{fig:shrink}
}
\end{figure}

We represent the conditional distribution of $R$ given $z$ by its three quartiles in the left panel of Figure \ref{fig:shrink}. We notice that considerable bias is already present at fairly small values of $z$. When $z=1.96$, the conditional median of $R$ is about 1.7.

\subsection{Coverage}
\begin{figure}[htp] \centering{
\includegraphics[scale=0.8]{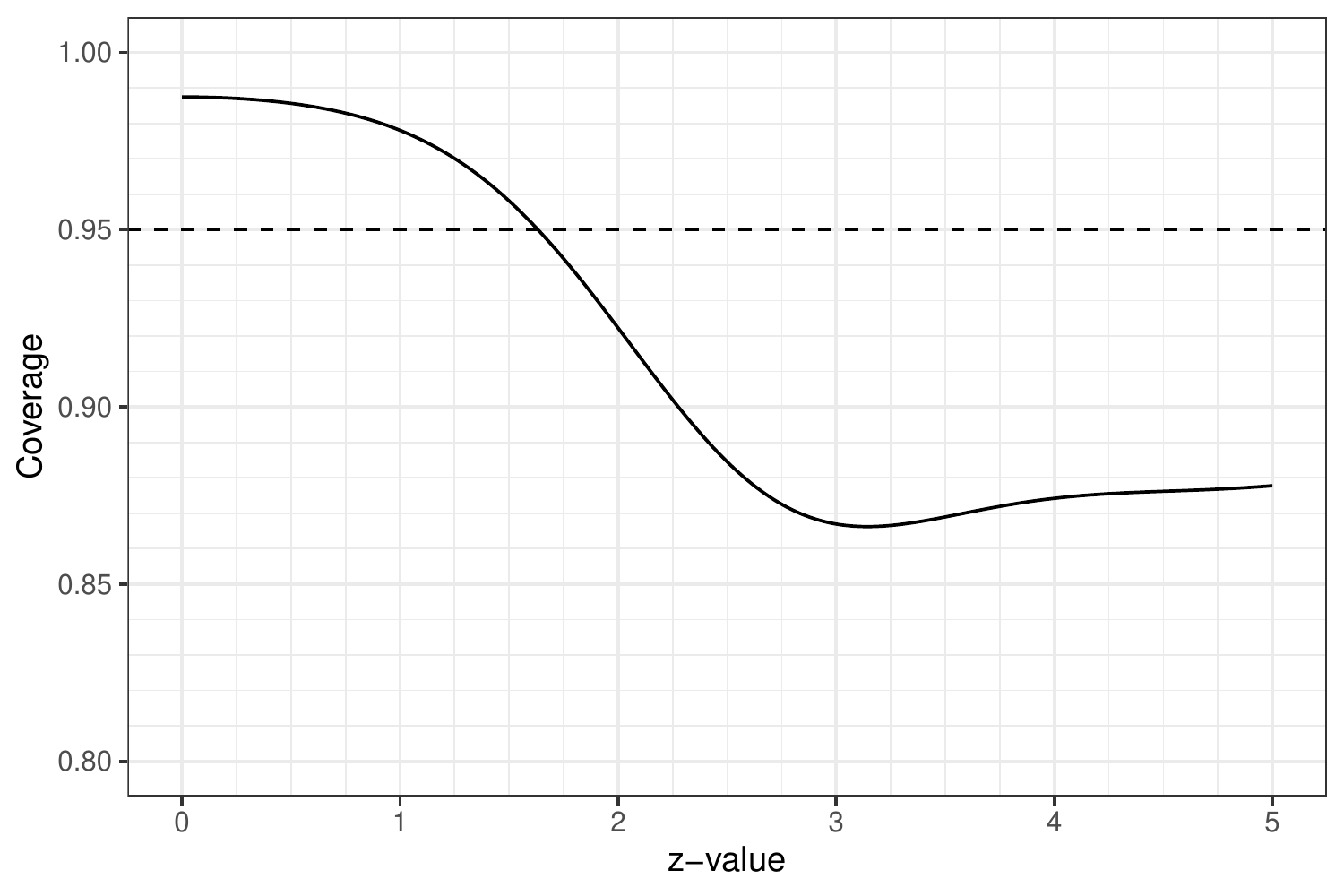}
\caption{Probability that the interval $[b-1.96\,s,b+1.96\,s]$ covers $\beta$ conditional on the $z$-value.}\label{fig:coverage}
}
\end{figure}

We conclude this section by considering the conditional coverage of the usual 95\% confidence interval given the observed $z$-value. Note that
\begin{equation}
P(b - 1.96\,s < \beta < b + 1.96\,s \mid z) = P(z - 1.96 < \SNR < z + 1.96 \mid z).
\end{equation}
So the conditional distribution of the SNR given $z$ (which we discussed in the previous sub-section) also yields the conditional coverage, which we plot in Figure \ref{fig:coverage}. We see that if $z$ exceeds 2, the conditional coverage is much lower than the nominal level of 95\%. 

\section{Shrinkage}\label{shrinkage}
Evidently, overestimation of the magnitude of the effect, and undercoverage of the confidence interval are very serious problems. In this section we propose a data-based approach to shrinkage.

As we discussed in the previous section, it is easy to compute the conditional distribution of the SNR given the $z$-value. This suggests an alternative estimator for $\beta$, namely
\begin{equation}
\hat \beta = s \E(\SNR \mid z).
\end{equation}
Since the (marginal) distribution of the SNR is a mixture of zero-mean normal distributions, $|\hat \beta|$ will be less than $|b|$. In other words, $\hat \beta$ is a shrinkage estimator. 

We can see in the left panel of Figure \ref{fig:shrink} that $|b|$ is typically very biased for $|\beta|$. In the right panel we show the three quantiles of the conditional distribution of $|\hat \beta|/|\beta|$ given $z$. We find that the shrinkage essentially resolves the winner's curse.

Since we have the conditional distribution of the SNR given $z$, we can construct an interval such that the conditional probability given $z$ of covering the SNR is 95\%. Scaling this interval by $s$ then yields a 95\% interval for $\beta$ which has the correct coverage conditionally on $z$.

Until now, we have conditioned on the $z$-value. However, we observe both $b$ and $s$, so we are ignoring some potentially relevant information. Now Assumption 1 makes it possible to estimate the complete joint distribution of $(\beta,b,s)$ from a sample of pairs $(b_1,s_1),\dots,(b_n,s_n)$. We are particularly interested in the conditional distribution of the effect $\beta$ given the observed estimate $b$ and its standard error $s$. Conditionally on $s$, Bayes' rule states
\begin{equation}
f(\beta \mid b,s) = \frac{f(\beta,b \mid s)}{f(b \mid s)} = \frac{f(b \mid \beta,s)f(\beta \mid s)} {\int f(b \mid \beta',s)f(\beta' \mid s) d \beta'}
\end{equation}
where $f(b \mid \beta,s)$ is the conditional density of $b$ given $\beta$ and  $s$, and $f(\beta \mid s)$ is the conditional density of $\beta$ given $s$. Since $f(b \mid \beta,s)$ is assumed to be known, we see that it is enough to have $f(\beta \mid s)$.  Now we can proceed by estimating $f(b \mid s)$ and then obtaining an estimate of $f(\beta \mid s)$ by deconvolution. Alternatively, we can make the following simplifying assumption.

\medskip \noindent {\bf Assumption 3} The standard error $s$ and the SNR $\beta/s$ are independent.

\medskip
This assumption  is very convenient, because it allows us to estimate the marginal distribution of the SNR, and then simply scale it by $s$ to obtain $f(\beta \mid s)$. In particular, we have
\begin{equation}
\E(\beta \mid b,s) = s\, \E(\SNR \mid z) = \hat \beta.
\end{equation}

So, while Assumption 3 is not necessary, it is certainly very convenient.  Essentially, it formalizes the notion that we can ignore any information in the pair $(b,s)$ that is not in the $z$-value.

Assumption 3 effectively means that we are using only background information about the distribution of the SNR. In fact, since Assumption 2 implies that the distribution of the SNR is symmetric around zero, we are really using only information about the absolute value of the SNR. Since there is a 1-1 correspondence between the absolute value of the SNR and the power, it is equivalent to say that we are using only information about the power.

Assumption 3 has one more practical advantage. Van Zwet and Gelman \cite{vanZwetGelman} show that Assumption 3 ensures that the inference about $\beta$ is equivariant under changes of the unit of measurement of the outcome. 

\section{Discussion}
We need many ways of looking at data. In this paper we offer a practical approach that is something of a frequentist-Bayesian hybrid in the hope that it may prove an interesting addition to whatever pure Bayesian or frequentist alternative a reader might prefer as their first choice of analysis based on their own philosophical preference. It is frequentist in the sense that it uses data collected on over 20,000 randomized controlled trials available in the Cochrane database. It is Bayesian in that it then supposes how one might use the distributional information provided by the database as a prior distribution to shrink the result from a given trial. Since we are estimating the prior information, our approach is perhaps best termed ``empirical Bayesian'' \cite{carlin2000empirical}. However, the very large sample size of over 20,000 means that we can safely ignore the uncertainty in our estimate of the prior.

We are not the first to use the Cochrane database to obtain prior information. Turner et al. \cite{turner2012predicting} and  Rhodes et al.\ \cite{rhodes2016implementing} used the database to get information about the heterogeneity of effects within a meta-analysis of a particular treatment. We do believe that we are the first to propose using the database to obtain prior information about the SNR, or equivalently, the power across all RCTs.

In offering our approach we are well aware that it may prove unacceptable to both Bayesians and frequentists for similar reasons. Both might reject the idea that a trial that one is about to analyze may be regarded as being exchangeable with the tens of thousands in the database. The frequentist may prefer to regard it as a realization of a theoretical infinity of trials having exactly the same property and the Bayesian may feel that there is much more to incorporating appropriate prior knowledge than simply using some very broad average. See \cite{spiegelhalter1994bayesian} for an extensive and balanced discussion of Bayesian approaches to RCTs.

Nevertheless, we feel that each may find the analysis of some interest. The Bayesian at least may find it calibrating. Given further information that suggests one ought to be optimistic it provides a lower bound for the probability of success and where the additional information encourages pessimism an upper one. The frequentist might consider that whereas such information is not usefully mixed with current information to form an estimate nevertheless, it may help to judge what future estimates from future trials might be.

In any case, we wish to stress, that we are not suggesting that this should replace other approaches but merely that it will provide further results that may be useful.


\end{document}